\documentclass[prd,nofootinbib,preprint,superscriptaddress]{revtex4}
\pdfoutput=1
\usepackage[T1]{fontenc}
\usepackage{amsmath,amssymb}
\usepackage{epsfig}
\usepackage{subfigure}
\usepackage{float}
\usepackage{graphicx}
\usepackage[usenames,dvipsnames]{color}
\usepackage{bigints}
\usepackage{slashed}
\usepackage{multirow}
\usepackage[colorlinks,citecolor=blue]{hyperref}
\usepackage{pdfpages}
\usepackage{color}
\usepackage{comment}
\usepackage[normalem]{ulem}

 \catcode`\@=11
 \def\lsim{\mathrel{\mathpalette\@versim<}}
 \def\gsim{\mathrel{\mathpalette\@versim>}}
 \def\@versim#1#2{\vcenter{\offinterlineskip
 \ialign{$\m@th#1\hfil##\hfil$\crcr#2\crcr\sim\crcr } }}
 \catcode`\@=12

 \catcode`@=12
\begin{document}
 \thispagestyle{empty}
 \begin{flushright}
 UCRHEP-T624\\
 Dec 2022\
 \end{flushright}
 \vspace{0.6in}
 \begin{center}
 {\LARGE \bf Dark $SU(2) \to Z_3 \times Z_2$ Gauge Symmetry\\}

 \vspace{1.5in}
{\bf Debasish Borah \\}
 \vspace{0.1in}
{\sl Department of Physics, Indian Institute of Technology
Guwahati, Assam 781039, India \\}
 
\vspace{0.2in}
 {\bf Ernest Ma\\}
 \vspace{0.1in}
{\sl Department of Physics and Astronomy,\\ 
University of California, Riverside, California 92521, USA\\}

 \vspace{0.2in}
{\bf Dibyendu Nanda \\}
 \vspace{0.1in}
{\sl School of Physics, Korea Institute for Advanced Study, Seoul 02455, Korea \\}

\end{center}
\vspace{1.0in}

\centerline{\bf Abstract}
\vspace{0.1in}
The dark sector is postulated to be invariant under an $SU(2)$ gauge 
symmetry, spontaneously broken by a Higgs quadruplet to a conserved 
residual $Z_3 \times Z_2$ symmetry.  The resulting dark matter 
phenomenology is studied.

\newpage
\baselineskip 24pt
\section{Introduction}
Whereas the stability of dark matter (DM)~\cite{Young:2016ala} is most 
likely due to an unbroken symmetry~\cite{Ma:2015xla, Ma:2019coj}, its nature 
and its origin are both unknown. It may well have something to do with a 
non-Abelian gauge symmetry, the simplest of which is 
$SU(2)$~\cite{Hambye:2008bq}.  The associated gauge bosons are 
presumably heavy from the spontaneous breaking of the gauge symmetry 
through Higgs scalars.  If a Higgs doublet is used, then the resulting 
theory has a residual $SU(2)$ global 
symmetry~\cite{Hambye:2008bq, Borah:2022phw}.  If both a doublet and a 
triplet are used, then the breaking may be gauge $SU(2)$ to gauge $U(1)$ to 
global $U(1)$~\cite{Ma:2018ouq}.  If other scalar multiplets 
are used, there are various possible outcomes~\cite{Etesi:1996urw}.  
In this paper, we study in detail the case of a Higgs $SU(2)$ 
quadruplet~\cite{Adulpravitchai:2009kd}, resulting in a dark discrete 
$Z_3$ symmetry~\cite{Agashe:2004ci,Ma:2007gq,Batell:2010bp}.  We show that 
an extra $Z_2$ symmetry emerges as the result of the structure of this 
particular scalar potential, which was not recognized previously.  We 
identify the possible dark matter 
candidates and study their phenomenology in relic abundance and direct 
detection.  We note that the idea of gauge $SU(2)$ family symmetry breaking to 
$A_4$ using a Higgs septuplet has also been studied~\cite{King:2018fke}.

\section{The Model}
Consider the origin of dark matter as coming from an $SU(2)$ gauge symmetry, 
broken by a Higgs quadruplet $\Phi$.  The complete scalar potential involving 
$\Phi$ is given by 
\begin{eqnarray}
V &=& -\mu^2 \Phi^\dagger \Phi + {1 \over 2} \lambda_1 (\Phi^\dagger \Phi)^2 - 
{10 \over 9} \lambda_2 (\Phi \tilde{\Phi})_3 (\Phi \tilde{\Phi})_3 \nonumber 
\\ &-& \left[ {5 \over 18} \lambda_3 (\Phi \Phi)_3 (\Phi \Phi)_3 + {5 \over 9} 
\lambda_4 (\Phi \Phi)_3 (\Phi \tilde{\Phi})_3 + {\rm h.c.} \right].
\end{eqnarray}
To justify the above, let $\Phi=(\phi_3,\phi_2,\phi_1,\phi_0)$, then the 
only quadratic invariant is $\Phi^\dagger \Phi$.  Now 
$(\Phi \Phi)_1 = (\Phi \Phi)_5 = 0$, whereas  
\begin{equation}
(\Phi \Phi)_3  = \begin{pmatrix}\sqrt{3 \over 10} \phi_3 \phi_1 - 
\sqrt{2 \over 5} \phi_2 \phi_2 + \sqrt{3 \over 10} \phi_1 \phi_3 \cr 
\sqrt{9 \over 20} \phi_3 \phi_0 - \sqrt{1 \over 20} \phi_2 \phi_1 - 
\sqrt{1 \over 20} \phi_1 \phi_2 + \sqrt{9 \over 20} \phi_0 \phi_3 \cr 
\sqrt{3 \over 10} \phi_2 \phi_0 - \sqrt{2 \over 5} \phi_1 \phi_1 + 
\sqrt{3 \over 10} \phi_0 \phi_2 \end{pmatrix},
\end{equation}
and
\begin{equation}
(\Phi \Phi)_3 (\Phi \Phi)_3 = {3 \over 5}(6 \phi_0 \phi_1 \phi_2 \phi_3 - 
3 \phi_0^2 \phi_3^2 + \phi_1^2 \phi_2^2) - {4 \sqrt{3} \over 5} 
(\phi_0 \phi_2^3 + \phi_1^3 \phi_3).
\end{equation}
As for $(\Phi \Phi)_7$, it may be neglected because 
$(\Phi \Phi)_7 (\Phi \Phi)_7 = -(\Phi \Phi)_3 (\Phi \Phi)_3$.

Consider now $\tilde{\Phi} = (\phi_0^*,-\phi_1^*,\phi_2^*,-\phi_3^*)$, which 
transforms as $\Phi$. 
\begin{equation}
(\Phi \tilde{\Phi})_3 = \begin{pmatrix}\sqrt{3 \over 10} \phi_3 \phi_2^* + 
\sqrt{2 \over 5} \phi_2 \phi_1^* + \sqrt{3 \over 10} \phi_1 \phi_0^* \cr 
-\sqrt{9 \over 20} \phi_3 \phi_3^* - \sqrt{1 \over 20} \phi_2 \phi_2^* + 
\sqrt{1 \over 20} \phi_1 \phi_1^* + \sqrt{9 \over 20} \phi_0 \phi_0^* \cr 
-\sqrt{3 \over 10} \phi_2 \phi_3^* - \sqrt{2 \over 5} \phi_1 \phi_2^* - 
\sqrt{3 \over 10} \phi_0 \phi_1^* \end{pmatrix}.
\end{equation}
Hence
\begin{eqnarray}
(\Phi \tilde{\Phi})_3 (\Phi \tilde{\Phi})_3 = -{1 \over 20} (3|\phi_0|^2  
+ |\phi_1|^2  - |\phi_2|^2  - 3|\phi_3|^2 )^2  
-  {1 \over 5} |\sqrt{3} \phi_0 \phi_1^* + 2\phi_1 \phi_2^* + \sqrt{3} \phi_2 
\phi_3^*|^2,
\end{eqnarray}
which is the same as $-(\Phi^\dagger \Phi)_3(\Phi^\dagger \Phi)_3$. 

Finally,
\begin{eqnarray}
(\Phi \Phi)_3 (\Phi \tilde{\Phi})_3 &=& {9 \over 10} (\phi_1 \phi_2 - 
\phi_0 \phi_3)(|\phi_0|^2-|\phi_3|^2) + {3 \over 10}(\phi_1 \phi_2 + 
3\phi_0 \phi_3)(|\phi_2|^2-|\phi_1|^2) \nonumber \\ 
&+& {\sqrt{3} \over 5} (\phi_2 \phi_2 \phi_2 \phi_3^* - 
\phi_0^* \phi_1 \phi_1 \phi_1 + 3\phi_0 \phi_1^* \phi_2 \phi_2 - 
3 \phi_1 \phi_1 \phi_2^* \phi_3),
\end{eqnarray}
which is the same as $-(\Phi \Phi)_7 (\Phi \tilde{\Phi})_7$.

As $\phi_0$ and $\phi_3$ develop vacuum expectation values, a residual $Z_3$ 
symmetry remains, maintained by $\phi_1 \sim \omega$, $\phi_2 \sim \omega^2$, 
with $\phi_{0,3} \sim 1$, where $\omega^3 = 1$.

Let $\langle \phi_{0,3} \rangle = v_{0,3}$, then $V$ is minimized with
\begin{eqnarray}
0 &=& v_0[-\mu^2 + \lambda_1(v_0^2+v_3^2)+\lambda_2(v_0^2-v_3^2)+\lambda_3 
v_3^2 + (3/2)\lambda_4 v_0 v_3]-\lambda_4 v_3^3/2, \\ 
0 &=& v_3[-\mu^2 + \lambda_1(v_0^2+v_3^2)+\lambda_2(v_3^2-v_0^2)+\lambda_3 
v_0^2 - (3/2)\lambda_4 v_0 v_3]+\lambda_4 v_0^3/2. 
\end{eqnarray}
Let $v_{D}=\sqrt{v_0^2+v_3^2}$, $c=\cos \theta = v_0/v_{D}$, 
$s=\sin \theta=v_3/v_D$, then
\begin{equation}
{\lambda_4 \over 2\lambda_2-\lambda_3} = {2sc(c^2-s^2) \over c^4+s^4-6s^2c^2} 
=  {1 \over 2} \tan 4\theta,
\label{eq9}
\end{equation}
\begin{equation}
v_{D}^2 = {2\mu^2 \over 2\lambda_1 + \lambda_3 + 
(\tan 4 \theta/2 \tan 2 \theta) (2\lambda_2-\lambda_3)}.
\end{equation}
The scalar masses and interactions are then functions of $v_D$, $\theta$, 
and $\lambda_{1,2,3}$. 

It is worth mentioning that the authors of \cite{Adulpravitchai:2009kd} have already shown that breaking $SU(2)$ symmetry with a quadruplet always leads to remnant $Z_3$ symmetry. In other words, it is always possible to redefine the $SU(2)$ basis to make $\langle \phi_1 \rangle =v_1 =0, \langle \phi_2 \rangle= v_2 = 0$, as chosen above. However, we have discovered that an extra accidental $Z_2$ symmetry appears from
its detailed structure.

\subsection{Dark Sector Mass Spectrum and Interactions}
The dark gauge bosons $X_{1,2,3}$ interact with $\Phi$ according to
\begin{equation}
\left|\partial \begin{pmatrix}\phi_3 \cr \phi_2 \cr \phi_1 \cr \phi_0 
\end{pmatrix} - 
ig_D \begin{pmatrix}{3 \over 2}X_3 & {\sqrt{3} \over 2}(X_1-iX_2) & 0 & 0 \cr 
{\sqrt{3} \over 2} (X_1+iX_2) & {1 \over 2} X_3 & X_1-iX_2 & 0 \cr 0 & 
X_1+iX_2 & -{1 \over 2} X_3 & {\sqrt{3} \over 2} (X_1-iX_2) \cr 0 & 0 & 
{\sqrt{3} \over 2} (X_1+iX_2) & -{3 \over 2} X_3 \end{pmatrix}
\begin{pmatrix}\phi_3 \cr \phi_2 \cr \phi_1 \cr \phi_0 \end{pmatrix} \right|^2.
\end{equation}

The masses of $X_{1,2,3}$ are 
\begin{equation}
M_{1,2}^2 = {3 \over 2} g_D^2 v_D^2, ~~~ M_3^2 = {9 \over 2} g_D^2 v_D^2,
\end{equation}
with $c \phi_1 - s \phi_2^*$ and $\sqrt{2} {\rm Im}(c \phi_0 - s \phi_3)$ as 
the longitudinal components of $(X_1 - i X_2)/\sqrt{2} \sim \omega$ and 
$X_3 \sim 1$ respectively under $Z_3$. 
The orthogonal components $\eta=s \phi_1 + c \phi_2^*$ and 
$\phi_I=\sqrt{2} {\rm Im}(s \phi_0 + c \phi_3)$ have masses 
\begin{equation}
M^2_\eta = {2 \over 3} \lambda_2 v_D^2 + {1 \over 2} M^2_{\phi_I}, 
~~~ M^2_{\phi_I} = -\left[2 \lambda_3 + {\tan 4 \theta \over  \tan 2 \theta} 
(2 \lambda_2 - \lambda_3) \right] v_D^2.
\end{equation}
The $2 \times 2$ mass-squared matrix spanning 
$\sqrt{2}{\rm Re}(\phi_0,\phi_3)$ is
\begin{equation}
M^2_R = \begin{pmatrix} 2(\lambda_1+\lambda_2)c^2 +(1/2)\lambda_4(3sc+s^3/c) 
& 2(\lambda_1-\lambda_2+\lambda_3)sc + (3/2)\lambda_4(c^2-s^2) \cr 
2(\lambda_1-\lambda_2+\lambda_3)sc + (3/2)\lambda_4(c^2-s^2) &  
2(\lambda_1+\lambda_2)s^2 -(1/2)\lambda_4(3sc+c^3/s) \end{pmatrix} v_D^2.
\end{equation}
Let $\zeta = \sqrt{2} {\rm Re} (c \phi_0 + s \phi_3)$ and 
$\phi_R = \sqrt{2} {\rm Re} (s \phi_0 - c \phi_3)$, then they are mass 
eigenstates, using the identity of Eq. \eqref{eq9}, with masses
\begin{eqnarray}
M^2_\zeta &=& [2\lambda_1 + \lambda_3 + {\cos^2 2 \theta \over \cos 4 \theta}
(2\lambda_2 - \lambda_3)]v_D^2, \\
M^2_{\phi_R} &=& {-1 \over \cos 4 \theta} 
(2\lambda_2 - \lambda_3) v_D^2.
\end{eqnarray}
Hence $\langle \zeta \rangle = \sqrt{2}v$ and $\langle \phi_R \rangle = 0$, 
with the latter (but not the former) connected to $\phi_I$ through $X_3$. 
This has the important consequence of the emergence of an extra $Z_2$ 
symmetry, under which $\phi_{R,I}$ are odd and $\zeta,\eta$ are even. 
As for the dark gauge bosons, $X_1 \pm iX_2$ are odd and $X_3$ is even.

The physical scalars are now contained in $\phi_{0,1,2,3}$ as follows.
\begin{eqnarray}
\phi_0 &\to& c v_D + {1 \over \sqrt{2}} (c\zeta + s\phi_R) + 
{is \over \sqrt{2}} \phi_I, \\  
\phi_1 &\to& s \eta, ~~~ \phi_2 ~\to~ c \eta^*, \\ 
\phi_3 &\to& s v_D + {1 \over \sqrt{2}} (s\zeta - c\phi_R) + 
{ic \over \sqrt{2}} \phi_I.
\end{eqnarray}

The gauge-scalar interactions are
\begin{eqnarray}
{\cal L}^{gs}_{int} &=& {3 \over 4}g_D X_3(\phi_R \partial \phi_I - 
\phi_I \partial \phi_R) + {i \over 2} g_D X_3 (\eta \partial \eta^* - 
\eta^* \partial \eta) \nonumber \\ &+& {\sqrt{3} \over 2} ig_D \left[
{X_1+iX_2 \over \sqrt{2}} (\phi_R -i\phi_I) \partial \eta - {X_1-iX_2 
\over \sqrt{2}} (\phi_R + i\phi_I) \partial \eta^* \right] \nonumber \\ 
&-& {\sqrt{3} \over 2} ig_D \left[{X_1+iX_2 \over \sqrt{2}} \eta \partial 
(\phi_R -i\phi_I) - {X_1-iX_2 \over \sqrt{2}} \eta^* \partial 
(\phi_R + i\phi_I) \right] \nonumber \\ 
&+& {3 \over 8}g_D^2(3X_3^2 + X_1^2 + X_2^2)(2\sqrt{2}v_D \zeta + \zeta^2 + 
\phi_R^2 + \phi_I^2) +{1 \over 4}g_D^2(X_3^2 + 7X_1^2 + 7X_2^2)|\eta|^2 
\nonumber \\ &-& \sqrt{3} g_D^2 X_3 \left( {X_1+iX_2 \over \sqrt{2}} \right)
(\phi_R-i\phi_I)\eta  -\sqrt{3} g_D^2 X_3 \left( {X_1-iX_2 \over \sqrt{2}} 
\right)(\phi_R+i\phi_I)\eta^* \nonumber \\ &+& \sqrt{3 \over 2} g_D^2 
(\sqrt{2}v_D + \zeta) \left[ \left( {X_1-iX_2 \over \sqrt{2}} 
\right)^2 \eta + \left( {X_1+iX_2 \over \sqrt{2}} \right)^2 \eta^* \right].
\end{eqnarray}

Let $x=\sin^2 2 \theta$, then
\begin{eqnarray}
M^2_{\phi_R} &=& {1 \over 2x-1}(2\lambda_2-\lambda_3)v_D^2, \\ 
M^2_{\phi_I} &=& -(2\lambda_2+\lambda_3)v_D^2 + M^2_{\phi_R}, \\ 
M^2_\eta &=& {2 \over 3} \lambda_2 v_D^2 + {1 \over 2} M^2_{\phi_I}, \\ 
M^2_\zeta &=& 2 \lambda_1 v_D^2 - {1 \over 2} M^2_{\phi_I} = 
2 \lambda_{D} v_D^2. 
\end{eqnarray}
Inverting the above,
\begin{eqnarray}
\lambda_1 v_D^2 &=& {1 \over 2} M^2_\zeta + {1 \over 4} M^2_{\phi_I}, \\ 
\lambda_2 v_D^2 &=& {3 \over 2} M^2_\eta - {3 \over 4} M^2_{\phi_I}, \\ 
\lambda_3 v_D^2 &=& M^2_{\phi_R} + {1 \over 2} M^2_{\phi_I} - 3 M^2_\eta, \\ 
x M^2_{\phi_R} &=& 3 M^2_\eta -  M^2_{\phi_I}\label{eq:mphi}.
\end{eqnarray}

\subsection{Dark Sector Interactions with the Standard Model}
With the addition of the Standard Model (SM) Higgs doublet $H$ which is 
a singlet under the dark gauge $SU(2)$, the complete scalar interactions 
are then given by
\begin{eqnarray}\nonumber
-{\cal L}_{int} &=& {M^2_\zeta \over 2v_D^2} \left[ {v_D \zeta^3 \over 
\sqrt{2}} + {1 \over 8} \zeta^4 + {1 \over 8} (\phi_R^2 + \phi_I^2)
(\phi_R^2 + \phi_I^2 + 4|\eta|^2) \right] + \left( {M^2_\zeta \over 2v_D^2} 
+ {M^2_{\phi_R} \over v_D^2} \right) \left[ {v_D \zeta \over \sqrt{2}} 
+ {1 \over 4} \zeta^2 \right] \phi_R^2 \\ 
&+& \left( {M^2_\zeta \over 2v_D^2} + {M^2_{\eta} \over v_D^2} \right) 
\left[ {v_D \zeta \over \sqrt{2}} + {1 \over 4} \zeta^2 \right] |\eta|^2 
+ \left( {M^2_\zeta \over 2v_D^2} + {3M^2_{\eta} \over v_D^2} -{M^2_{\phi_R} 
\over v_D^2} \right) \left[ {v_D \zeta \over \sqrt{2}} + {1 \over 4} \zeta^2 
\right] \phi_I^2 \nonumber \\ 
&+& \left( {M^2_\zeta \over 4v_D^2} + {M^2_{\eta} \over 3v_D^2}  \right) 
|\eta|^4 + {1 \over 3\sqrt{3}}\left( {2M^2_{\phi_R} \over v_D^2} - 
{3M^2_{\eta} \over v_D^2}  \right) \left[ v_D + {\zeta \over \sqrt{2}} \right] 
(\eta^3 + {\eta^*}^3) \nonumber \\
&+& \lambda_{H\phi} \left(H^\dagger H \right)\left( \Phi^\dagger \Phi \right) 
- \mu_{H}^2 \left(H^\dagger H \right) +\frac{\lambda_{H}}{2} 
\left(H^\dagger H \right)^2,
\label{scalarint}
\end{eqnarray}

where
\begin{eqnarray}
 (H^\dagger H) &=& \left(v_H + \frac{h}{\sqrt{2}}\right)^2,\\
 (\Phi^\dagger \Phi ) &=& \left(v_D + \frac{\zeta}{\sqrt{2}}\right)^2 + 
\frac{1}{2} \phi_R^2 + \frac{1}{2} \phi_I^2 + |\eta|^2.
\end{eqnarray}
From the above, it is clear that the scalar portal coupling $\lambda_{H\phi}$ 
is the only connection between the dark and SM sectors.  It allows $h$ to 
couple to $|\eta|^2$, $\phi^2_{R,I}$, as well as $\zeta^2$, and leads to the 
mixing between $h$ and $\zeta$, with mass-squared matrix given by
\begin{equation}
\mathcal{M}_{h\zeta}^2=\begin{pmatrix}2 \lambda_H v_{H}^2 & 2 \lambda_{H\phi} 
v_H v_D \cr 2 \lambda_{H\phi} v_H v_D & 2 \lambda_D v_{D}^2 \end{pmatrix}.
\label{massmatrix}
\end{equation}
This mixing also implies that the dark gauge bosons $X_{1,2,3}$ are also 
coupled to the SM through $\lambda_{H\phi}$. 

The $\lambda_{H\phi}$ coupling allows the $Z_3 \times Z_2$ dark matter to 
scatter off nucleons through the SM Higgs boson.  It is required to be very 
small, of order $10^{-4}$, to agree with present experimental data. However, 
it should also be big enough so that the physical scalar which is mostly 
$\zeta$ could decay rapidly through its small $h$ component to SM particles. 
For the purpose of discussing the relic abundance, the role of $\lambda_{H\phi}$ can remain sub-dominant, as in our previous study~\cite{Borah:2022phw}.  Under the residual 
dark $Z_3 \times Z_2$ symmetry, the SM particles as well as $\zeta$ and $X_3$ 
do not transform, whereas 
\begin{equation}
\eta \sim (\omega,+), ~~~ \phi_{R,I} \sim (1,-), ~~~ (X_1 - iX_2)/\sqrt{2} 
\sim (\omega,-).
\end{equation}
This means that there are at least two stable dark-matter components.  
We will assume that $\eta$ is the lightest, and $\phi_I$ the second 
lightest, with both annihilating to $\zeta$.  We choose our input 
parameters to be
$$ g_D,v_D,x, \lambda_{H \phi}, M_{\zeta},M_{\eta}, M_{\phi_I},$$
with the mass hierarchy
\begin{equation}
M_{\phi_R} > M_{X_3} > M_{X_{1,2}} > M_{\phi_I} > M_\eta > M_\zeta.
\label{eq:hierarchy}
\end{equation}
Since the dark sector particles except $\zeta$ decays within the dark sector itself, it is also possible to have three-component DM by choosing a compressed mass spectrum, as studied in \cite{Choi:2021yps} in the context of $U(1)_X \to Z_3 \times Z_2$ dark matter. Unlike this work containing several scalar fields and hence more free parameters, we have very limited number of free parameters due to the presence of only one scalar multiplet responsible for dark $SU(2)$ gauge symmetry breaking and resulting DM phenomenology. We discuss the details of two-component DM phenomenology in the upcoming section. 

\begin{figure}[hbt!]
    \centering
        \includegraphics[scale=0.35]{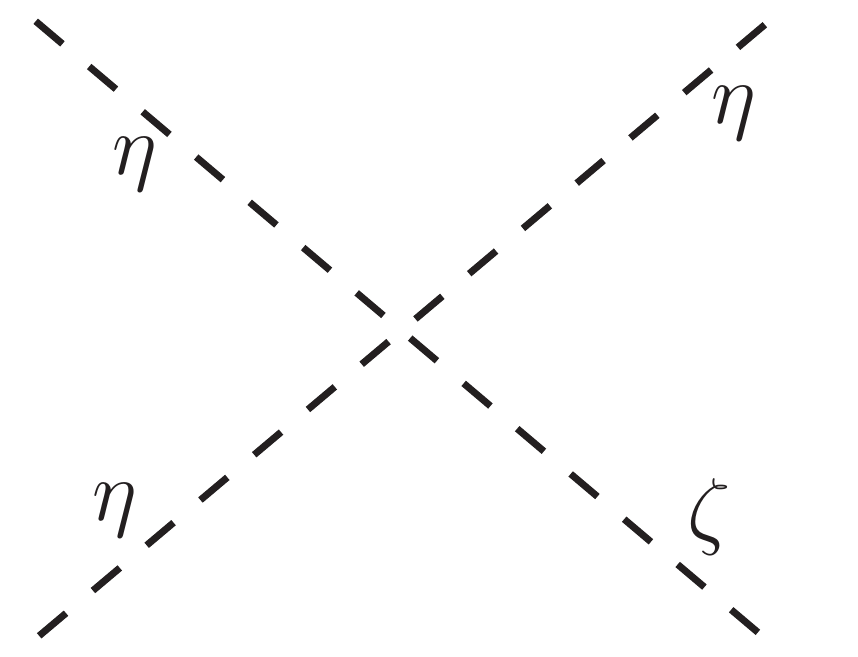}\,
                \includegraphics[scale=0.35]{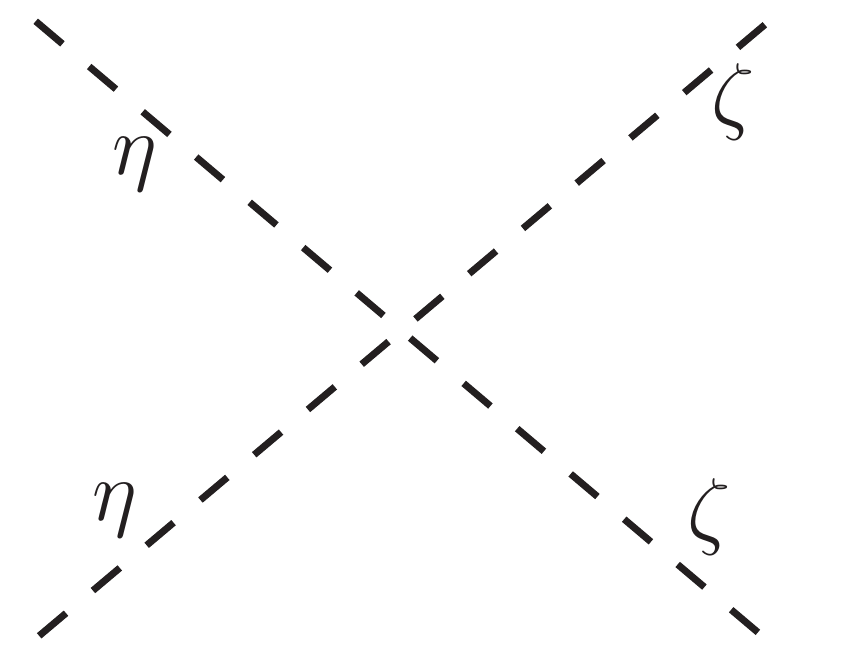} \,
                \includegraphics[scale=0.35]{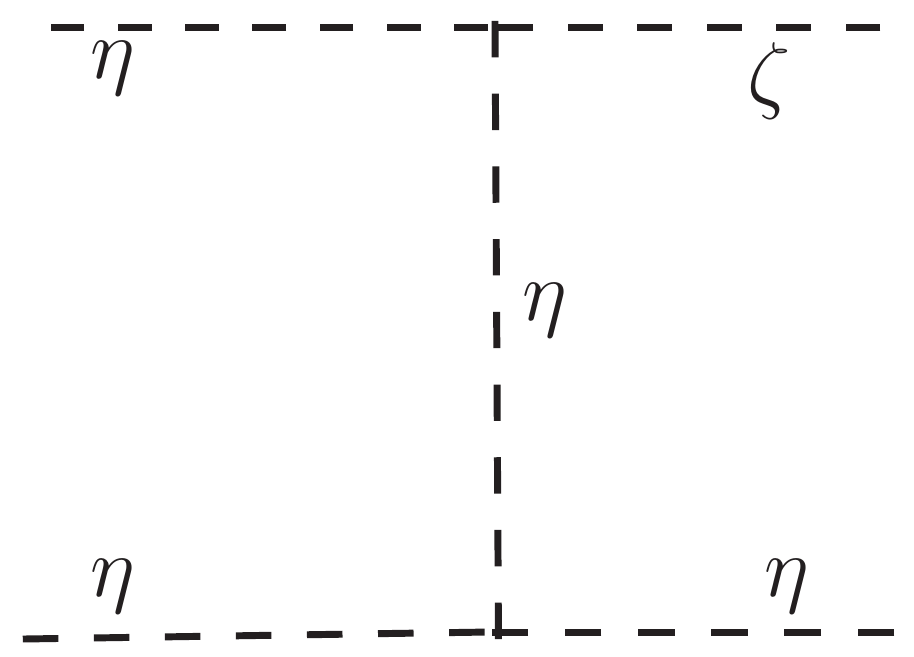}\,
                      \includegraphics[scale=0.35]{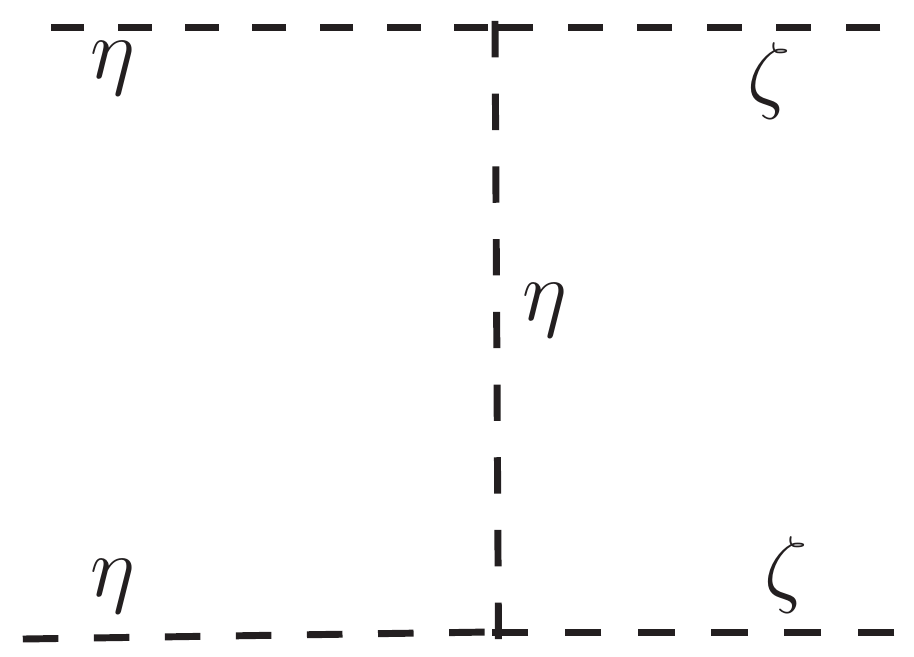}\\  \includegraphics[scale=0.35]{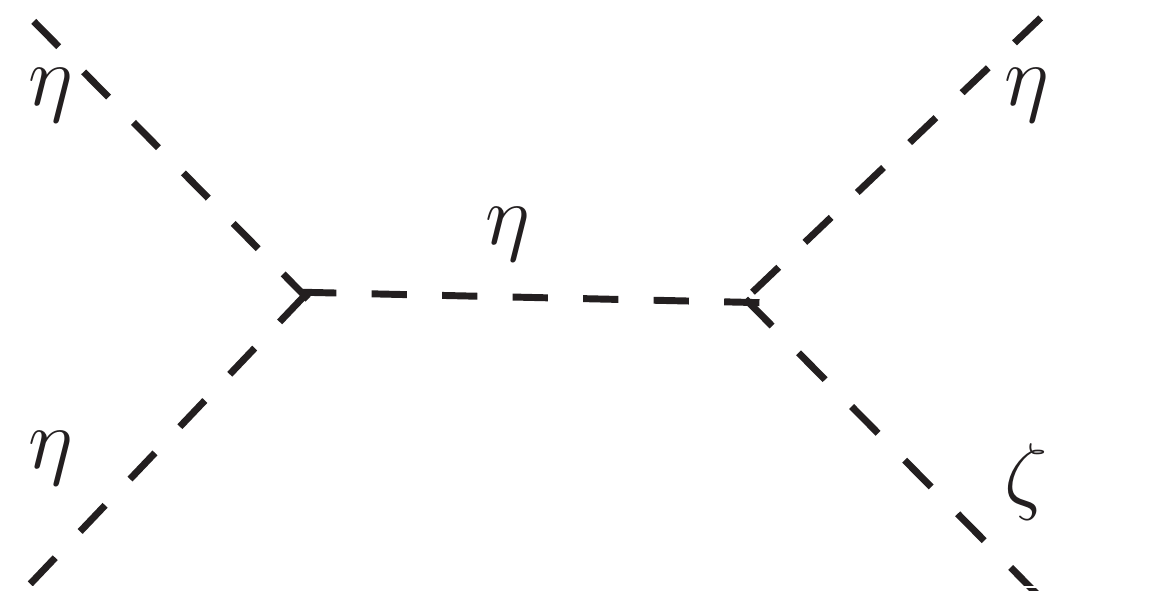}\,
                \includegraphics[scale=0.35]{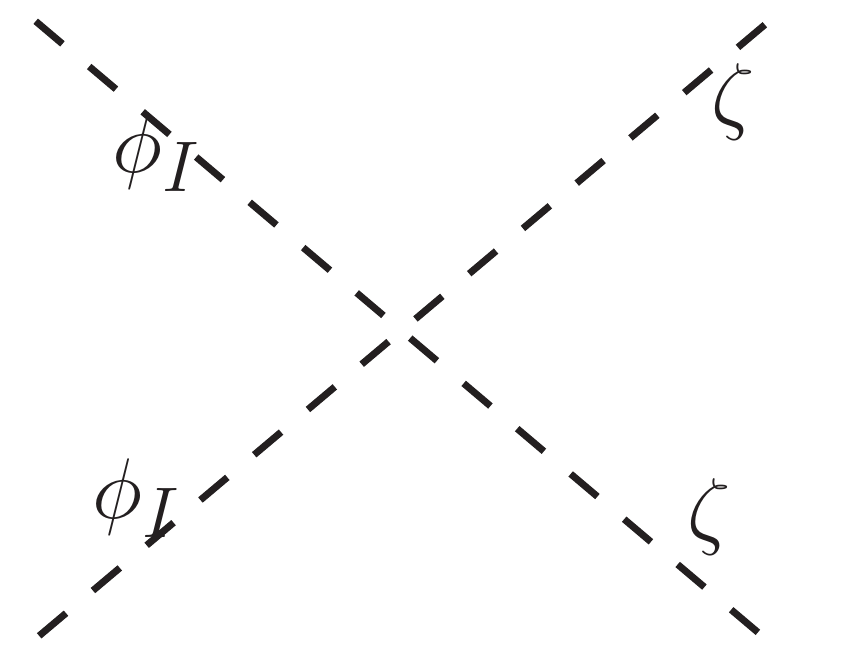}\,
                \includegraphics[scale=0.35]{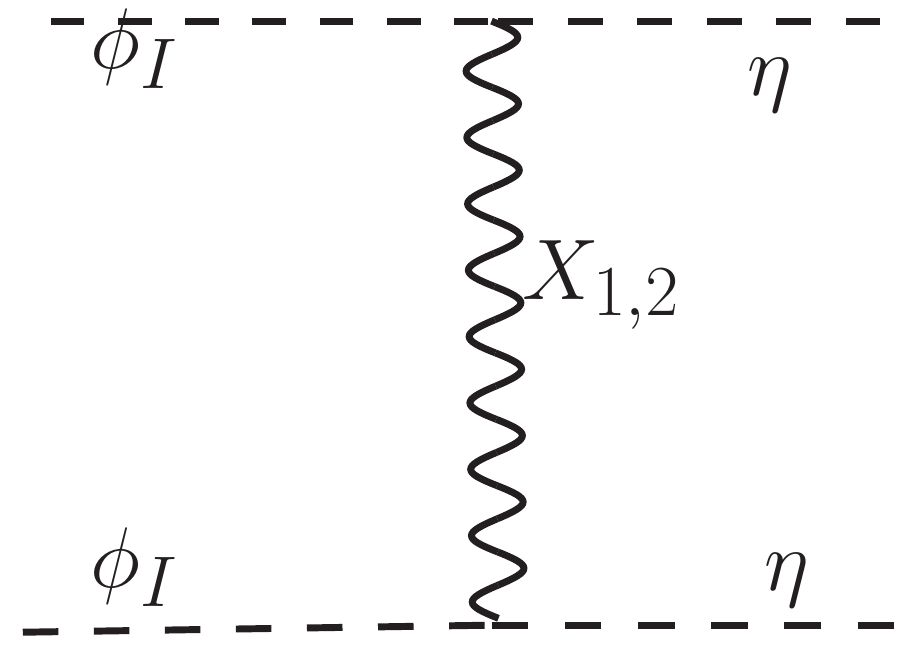}\,
                \includegraphics[scale=0.35]{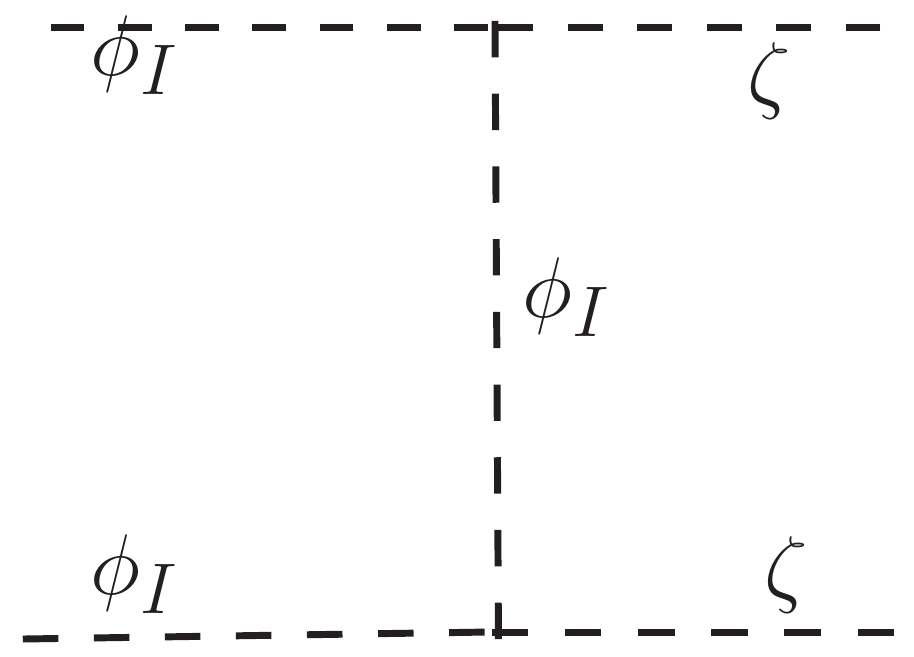}
                \caption{Dominant Semi-annihilation and annihilation channels for DM components $\eta$ and $\phi_I$. }
    \label{fig:feyn}
\end{figure}

\section{Dark matter phenomenology}
As mentioned above, we have a natural two-component DM model due to the 
residual $Z_3 \times Z_2$ discrete symmetry from the spontaneous breaking 
of dark gauge $SU(2)$ by a Higgs quadruplet.  To discuss the relic density 
of two-component dark matter \cite{Cao:2007fy, Zurek:2008qg, Liu:2011aa, Belanger:2011ww, Adulpravitchai:2011ei} quantitatively, we first write the coupled Boltzmann equations (BEs) for comoving number densities of the two DM candidates which we assume to be $\eta$ and $\phi_I$ respectively. We have assumed that $\eta$ is the lightest $Z_3$ odd particle in the spectrum and is absolutely stable. On the other hand, due to the unbroken $Z_2$ symmetry, $\phi_I$ is also stable along with $\eta$. In Fig. \ref{fig:feyn}, we show the dominant Feynman diagrams contributing to the annihilation and semi-annihilation processes of both $\eta$ and $\phi_I$. Since $\phi_I$ is neutral under $Z_3$ it gives rise to typical annihilation channels into lighter particles while $\eta$, being charged under $Z_3$, also has semi-annihilation processes like $\eta \eta \rightarrow \eta \zeta$, typical of $Z_3$ dark matter. We have assumed that $X_{1,2,3}$, $\phi_R$ are heavier and hence these final states will not contribute to the relic density calculation. Also, as mentioned earlier, the Higgs portal coupling $\lambda_{H \phi}$ is assumed to be small so that the annihilation processes to SM final states remain sub-dominant. However, this coupling cannot be arbitrarily small as the dark sector gets thermalised with the SM by virtue of Higgs portal interactions only. This leads to a lower bound on $\lambda_{H \phi}$ from the requirement of dark sector thermalisation. Considering the point-like interaction $\zeta \zeta \rightarrow h h$ with a cross-section $\sigma \approx \lambda^2_{H \phi}/(32 \pi s)$, one can find the lower bound on $\lambda_{H \phi}$ by demanding the rate of interaction to be greater than the Hubble expansion rate at some high-temperature T. This leads to $ n^{\rm eq} \langle \sigma v \rangle (T) > {\bf H}(T) = 1.66 \sqrt{g_*} T^2/M_{\rm Pl}$. Demanding the dark sector to thermalise with the SM above the TeV scale $(T \geq 1 \, {\rm TeV})$ leads to $\lambda_{H \phi} \gtrsim 10^{-7}$. Choosing it to be very small, however, can lead to very early DM-SM decoupling. This requires the tracking of dark sector temperature evolution till the epoch of dark sector freeze-out. In this work, we consider $\lambda_{H \phi}$ to be in the ballpark of $10^{-4}-10^{-3}$ and assume the dark sector and the SM baths to evolve with a common temperature.

Choosing the variable to be $x_i=(M_i/T)$ with $i= 1,2$ for $\eta, \phi_I$ respectively, the BEs in the limit of two stable dark matter candidates with the possible annihilations specified above, can be written as
\begin{eqnarray}\nonumber
\frac{dY_{\eta}}{dx_1} &=& \frac{\beta s}{{\bf H} x_1}\bigg(-\left<\sigma (\eta \eta \rightarrow {\rm SM \, SM})v_{\rm rel} \right>\left[Y_{\eta}^2- \left(Y_{\eta}^{\rm eq}\right)^2 \right] - \left<\sigma (\eta \eta \rightarrow \zeta \zeta)v_{\rm rel} \right>\left[Y_{\eta}^2- (Y_{\eta}^{\rm eq})^2 \right] \nonumber \\ && -\left<\sigma (\eta \eta \rightarrow \eta h)v_{\rm rel} \right>\left[Y_{\eta}^2- Y_{\eta}^{\rm eq}Y_{\eta} \right] -\left<\sigma (\eta \eta \rightarrow \eta \zeta)v_{\rm rel} \right>\left[Y_{\eta}^2- Y_{\eta}^{\rm eq}Y_{\eta} \right] \nonumber \\ &&
+ \left<\sigma (\phi_I \phi_I \rightarrow \eta \eta)v_{\rm rel} \right>\left[Y_{\phi_I}^2 - \frac{\left(Y_{\phi_I}^{\rm eq}\right)^2}{\left(Y_{\eta}^{\rm eq}\right)^2}Y_{\eta}^2 \right]\bigg),\label{BE:eta:Y}\\
\frac{dY_{\phi_I}}{dx_2} &=& \frac{\beta s}{{\bf H} x_2} \Bigg(-\left<\sigma (\phi_I \phi_I \rightarrow {\rm  SM \, SM})v_{\rm rel} \right>\left[Y_{\phi_I}^2- \left(Y_{\phi_I}^{\rm eq}\right)^2 \right] -\left<\sigma (\phi{_I} \phi_{I} \rightarrow \zeta \zeta)v_{\rm rel} \right>\left[Y_{\phi_{I}}^2- (Y_{\phi_{I}}^{\rm eq})^2 \right] \nonumber \\ && - \left<\sigma (\phi_I \phi_I \rightarrow \eta \eta)v_{\rm rel} \right>\left[Y_{\phi_I}^2 - \frac{\left(Y_{\phi_I}^{\rm eq}\right)^2}{\left(Y_{\eta}^{\rm eq}\right)^2}Y_{\eta}^2 \right]\Bigg),
\label{BE:phiI:Y}
\end{eqnarray}
where $Y_x=(n_x/s)$ denotes comoving number density for $x \equiv \eta, \phi_I$ with 
$$\beta (T)= 1 + \frac{1}{3} \frac{T}{g_s(T)}\frac{d g_s(T)}{dT}$$ 
and ${\bf H}$ being the Hubble parameter. The first term of the right hand side of Eq. \eqref{BE:eta:Y} shows the annihilation to the SM final states whereas the second, third and fourth terms show the annihilation to $\zeta \zeta$, semi-annihilation to Higgs and $\zeta$ respectively. The last term represents the conversion processes between the two dark matter particles. Similarly, the first term of the right hand side of Eq. \eqref{BE:phiI:Y} is for the annihilation to the SM final states and the second and the third terms are for the annihilation to $\zeta$ and conversion to $\eta$ respectively. Among all these processes the two $\zeta$ final state and the semi-annihilation processes give the dominant contribution to the relic density due to the chosen mass hierarchy as well as small Higgs portal interactions. In order to calculate the thermally averaged annihilation cross-sections and solve the above BEs numerically, we use \texttt{micrOMEGAs} \cite{Belanger:2014vza}.

\begin{figure}[hbt!]
    \centering
    \includegraphics[scale=0.45]{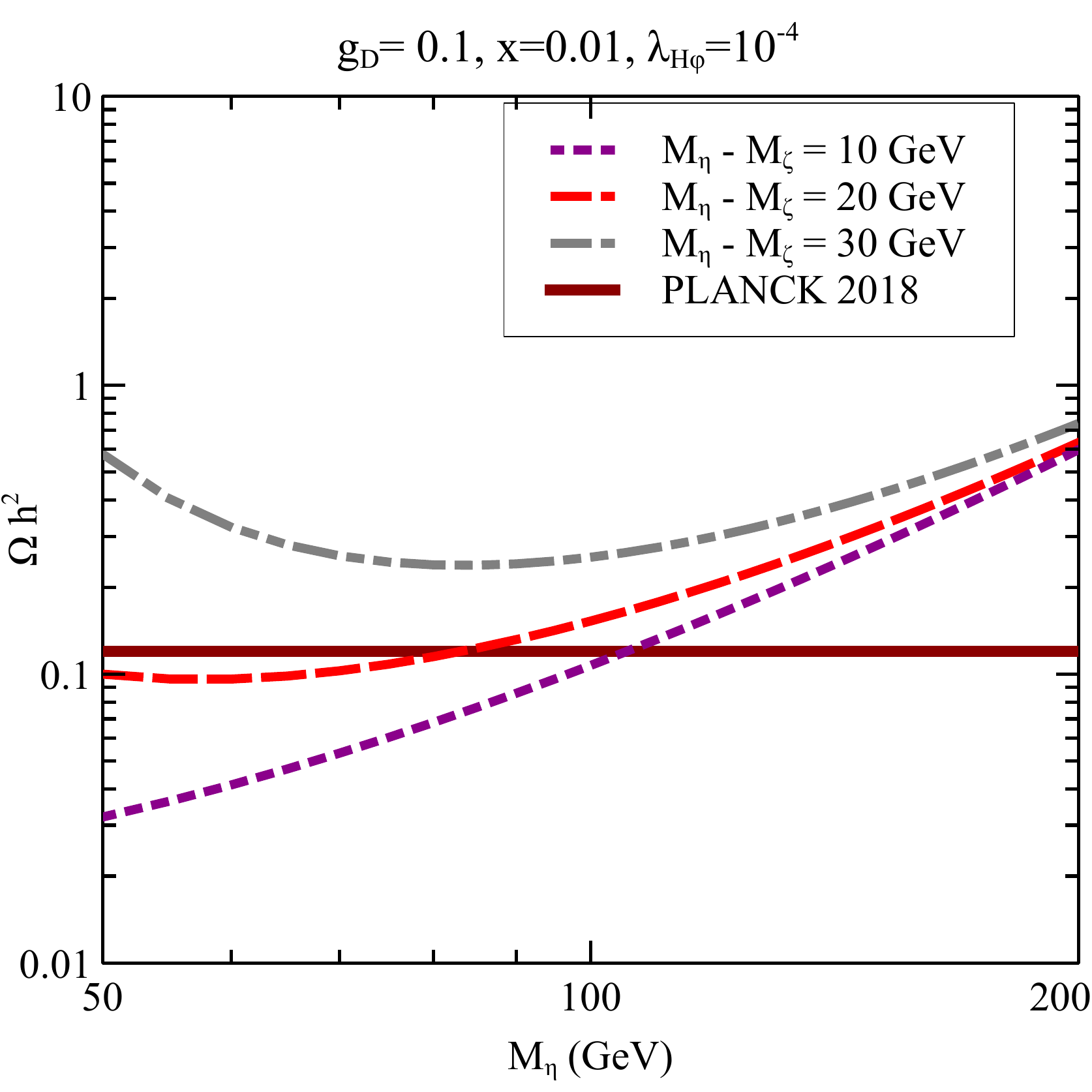}
    \caption{Total relic density of dark matter as a function of $M_{\eta}$ for three different mass splitting between $M_{\eta}$ and $M_{\zeta}$.}
    \label{fig:line}
\end{figure}

\begin{figure}[hbt!]
    \centering
    \includegraphics[scale=0.45]{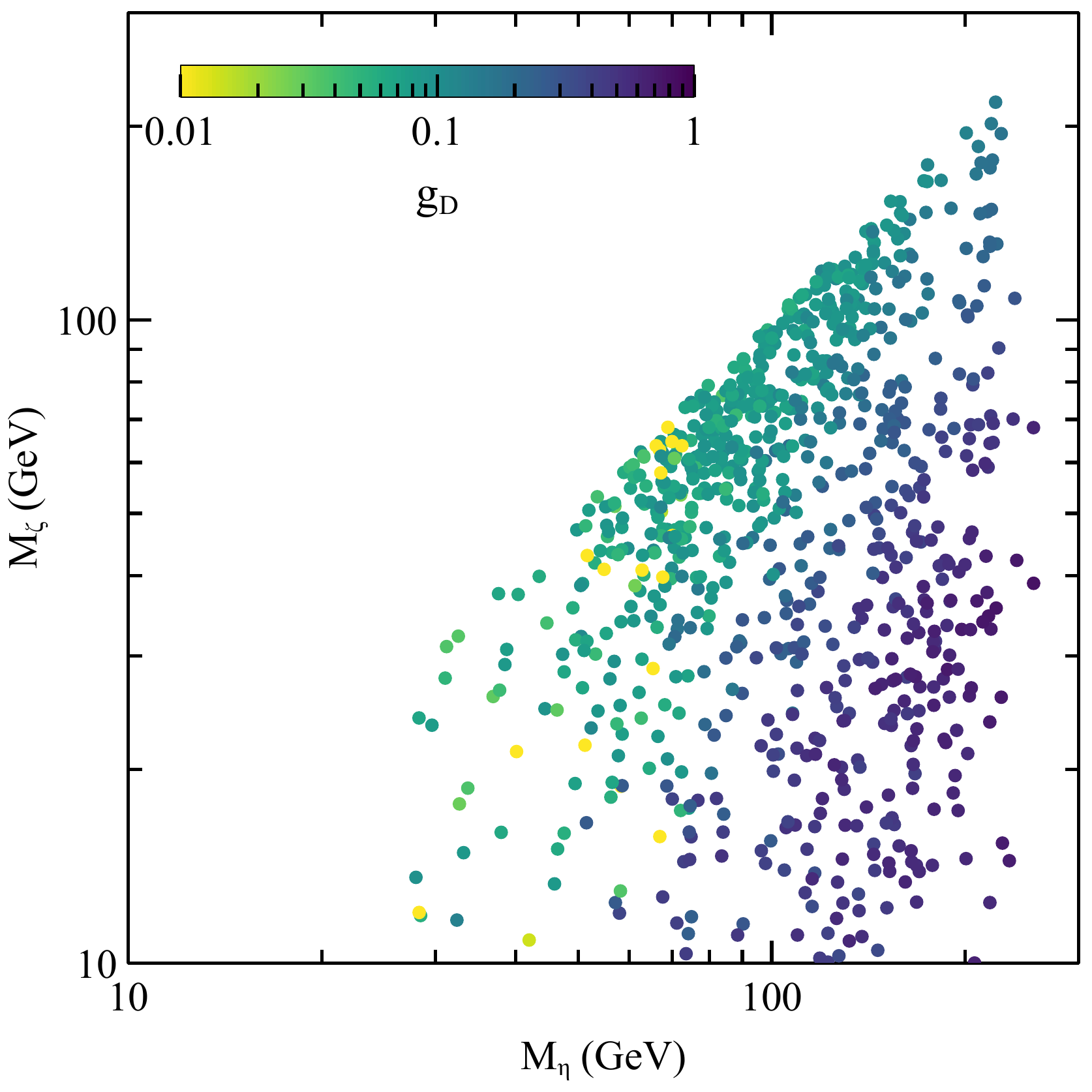}\,
        \includegraphics[scale=0.45]{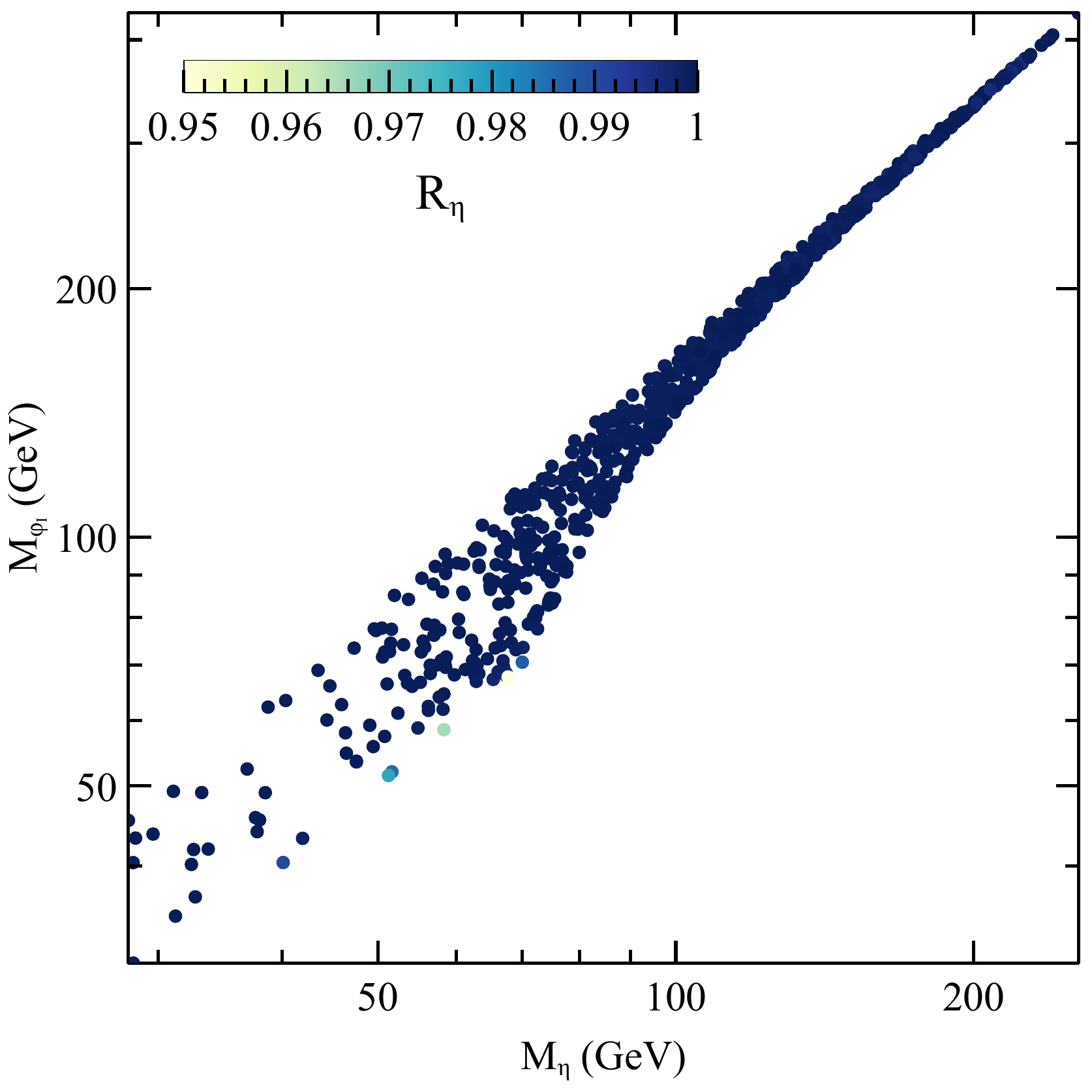}
    \caption{Left panel: Allowed parameter space in $M_{\eta}-M_{\zeta}$ plane from total DM relic abundance criteria. Right panel: Allowed parameter space in $M_{\eta}-M_{\phi_I}$ plane from total DM relic abundance criteria. Color bar in the left (right) panel shows the variation of dark gauge coupling $g_D$ (relative relic abundance of $\eta$ given by  $R_{\eta} = \Omega_{\eta}/(\Omega_{\eta}+\Omega_{\phi_I})$).}
    \label{fig:2comp}
\end{figure}

\begin{figure}[hbt!]
    \centering
    \includegraphics[scale=0.45]{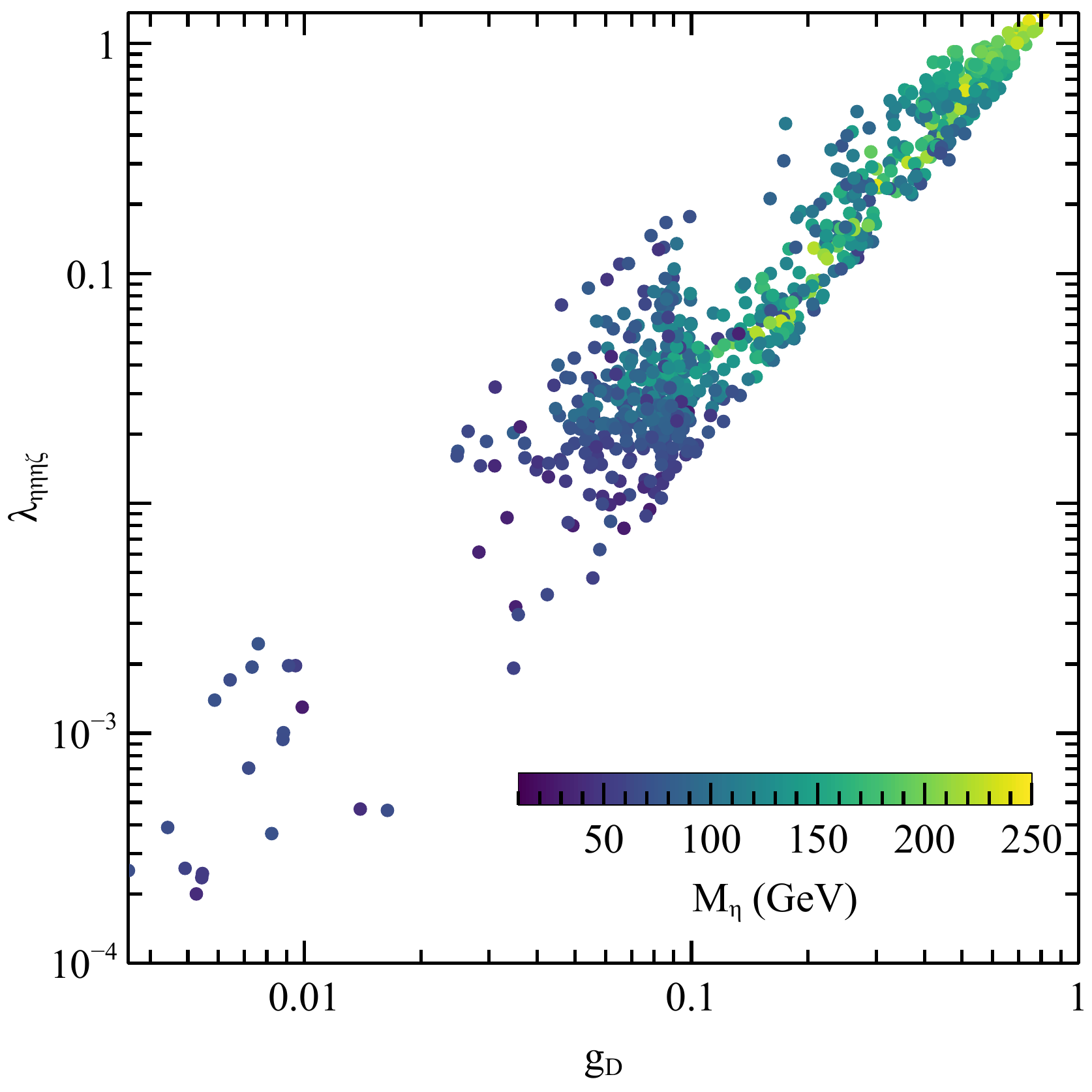}\,
        \includegraphics[scale=0.45]{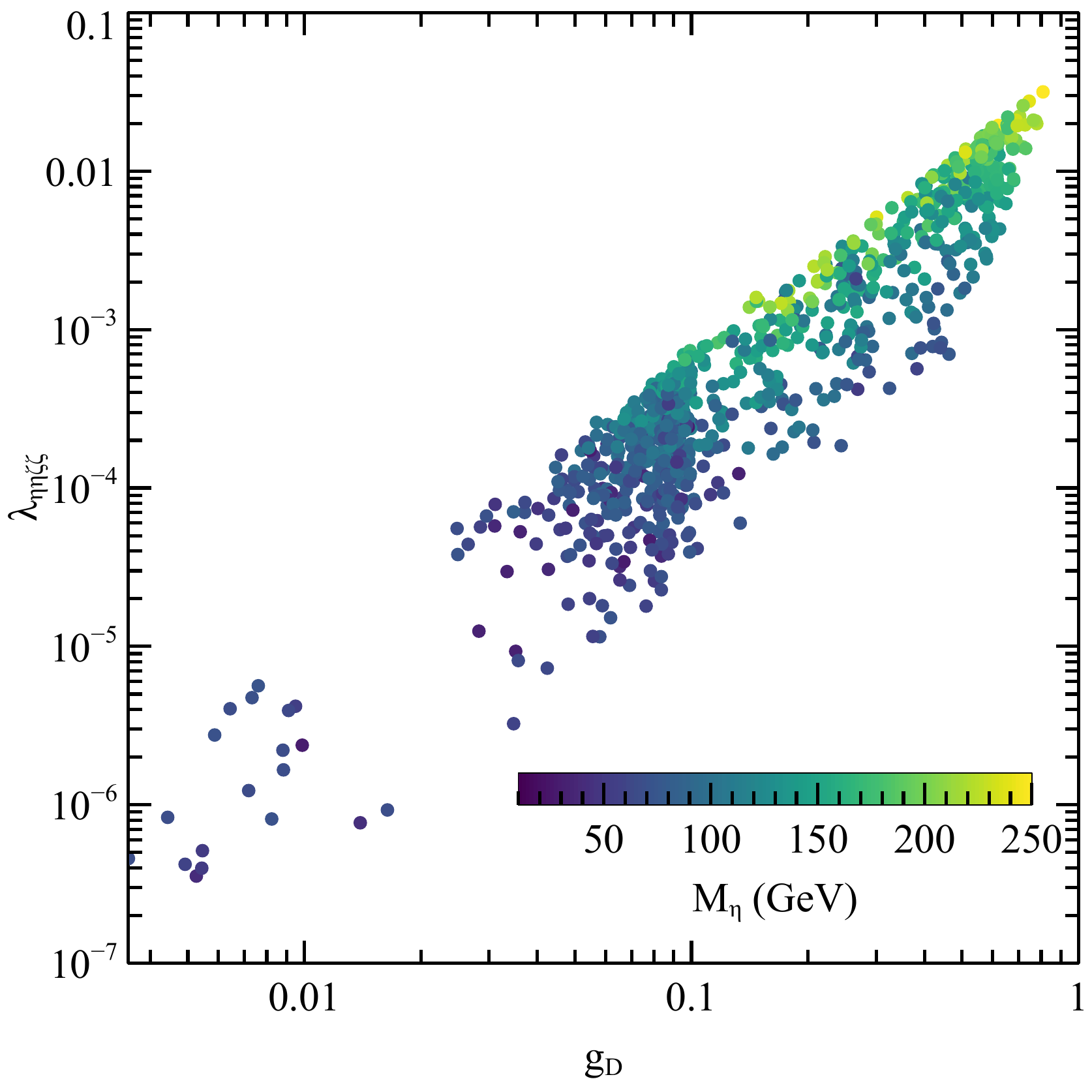}
    \caption{Allowed parameter space in $g_D-\lambda_{\eta\eta\eta\zeta}$ plane (Left panel) and in $g_D-\lambda_{\eta\eta\zeta\zeta}$ plane from total DM relic abundance criteria.}
    \label{fig:scalar_vcoup}
\end{figure}

Let us discuss the numerical results of the model. As discussed above, we have two different components of stable dark matter in this scenario, the lightest ${Z}_3$-charged particle $\eta$ and the lightest $Z_2$-odd particle $\phi_I$. However, in this study we have restricted ourselves to the regime where $M_{\eta}$ is smaller than $M_{\phi_I}$. For our analysis, we have considered the mixing between the dark sector and the SM to be small governed by the smallness of $\lambda_{H \phi}$ which is also motivated by the direct detection constrain from various direct detection experiments. We consider the two DM masses to be free parameters with $M_{\eta} < M_{\phi_I}$. We also choose the remaining free parameters $g_D, v_D, x, \lambda_{H \phi}, M_{\zeta}$ in a way which maintains the dark sector mass hierarchy in Eq. \eqref{eq:hierarchy}. In Fig. \ref{fig:line}, we have shown the total DM relic density as function of $M_{\eta}$ for three different benchmark values of $M_{\zeta}$. 
One can note that total relic density decreases with decreasing mass splitting between $M_{\eta}$ and $M_{\zeta}$. For a fixed value of $M_{\eta}$, as we increase $M_{\zeta}$ in order to decrease the mass splitting $M_\eta-M_\zeta$, it increases the scalar interactions of $\eta$ as seen from Eq. \eqref{scalarint} enhancing its annihilation rate thereby decreasing the relic abundance. Due to such dependence of scalar couplings on $M_{\zeta}$, $\eta \eta \rightarrow \eta \zeta$ and $\eta \eta \rightarrow \zeta \zeta$ processes dominate the relic abundance. Notably, for the chosen mass hierarchy of $X_{1,2}, \phi_R, \eta$, the total DM relic is dominated by $\eta$ as strong annihilation and conversion rates of heavier DM candidate $\phi_I$ reduce its relative abundance. In order to make it clear, we scan the parameter space for the chosen hierarchy. In the left panel of Fig. \ref{fig:2comp}, we have shown the allowed parameter space in $M_{\eta} - M_{\zeta}$ plane from the observed total relic density constraints of dark matter. The right panel of the same figure shows the allowed parameter space in $M_{\eta} - M_{\phi_I}$ plane. Color bar in the left panel shows the variation of the gauge coupling $g_D$ whereas the color bar in the right panel shows the variation of relative relic abundance of $\eta$ $(R_{\eta} = \Omega_{\eta}/(\Omega_{\eta}+\Omega_{\phi_I}))$. As seen from the right panel plot, the total relic density is dominated by $\eta$ for the chosen mass hierarchy. This is due to large annihilation rate of $\phi_I$ by virtue of its large coupling with $\zeta$ for the chosen mass spectrum of different scalars, as seen from the scalar interactions in Eq. \eqref{scalarint}. Apart from the large annihilation rate, we have also set $M_{\phi_I}$ to be heavier than $M_{\eta}$ which gives more Boltzmann suppression in the number density of $\phi_I$. This can be seen that $\phi_I$ contributes nearly $5\%$ of the total relic abundance when $M_{\phi_I}\sim M_{\eta}$.  In Fig. \ref{fig:scalar_vcoup}, we have shown the variation of effective scalar couplings $\lambda_{\eta \eta \eta \zeta}, \lambda_{\eta \eta \zeta \zeta}$ as a function of $g_D$ where the color bar represents the variation of $M_{\eta}$. These couplings are not free parameters and as shown in Eq. \eqref{scalarint}, they depend upon other parameters as
\begin{equation}
   \lambda_{\eta \eta \eta \zeta}={1 \over 3\sqrt{6}}\left( {2M^2_{\phi_R} \over v_D^2} - 
{3M^2_{\eta} \over v_D^2}  \right),\,\, \lambda_{\eta \eta \zeta \zeta} = \frac{1}{4} \left( {M^2_\zeta \over 2v_D^2} + {M^2_{\eta} \over v_D^2} \right).
\end{equation}
The correlations between $g_D$, $M_{\eta}$ and the quartic couplings shown in Fig. \ref{fig:scalar_vcoup} can be understood as follows. Increasing $g_{D}$ increases the $\lambda_{\eta\eta\eta\zeta}$ and $\lambda_{\eta\eta\zeta\zeta}$ as shown in the figure and this also increases the annihilation cross section of $\eta$. However, this was compensated by increasing $M_{\eta}$ as for point interactions $\sigma\sim \frac{\lambda^2}{8\pi M_{\eta}^2}$. Similar correlation can also be seen between $g_D$ and $M_{\eta}$ in Fig. \ref{fig:2comp}.

\begin{figure}
    \centering
\includegraphics[scale=0.5]{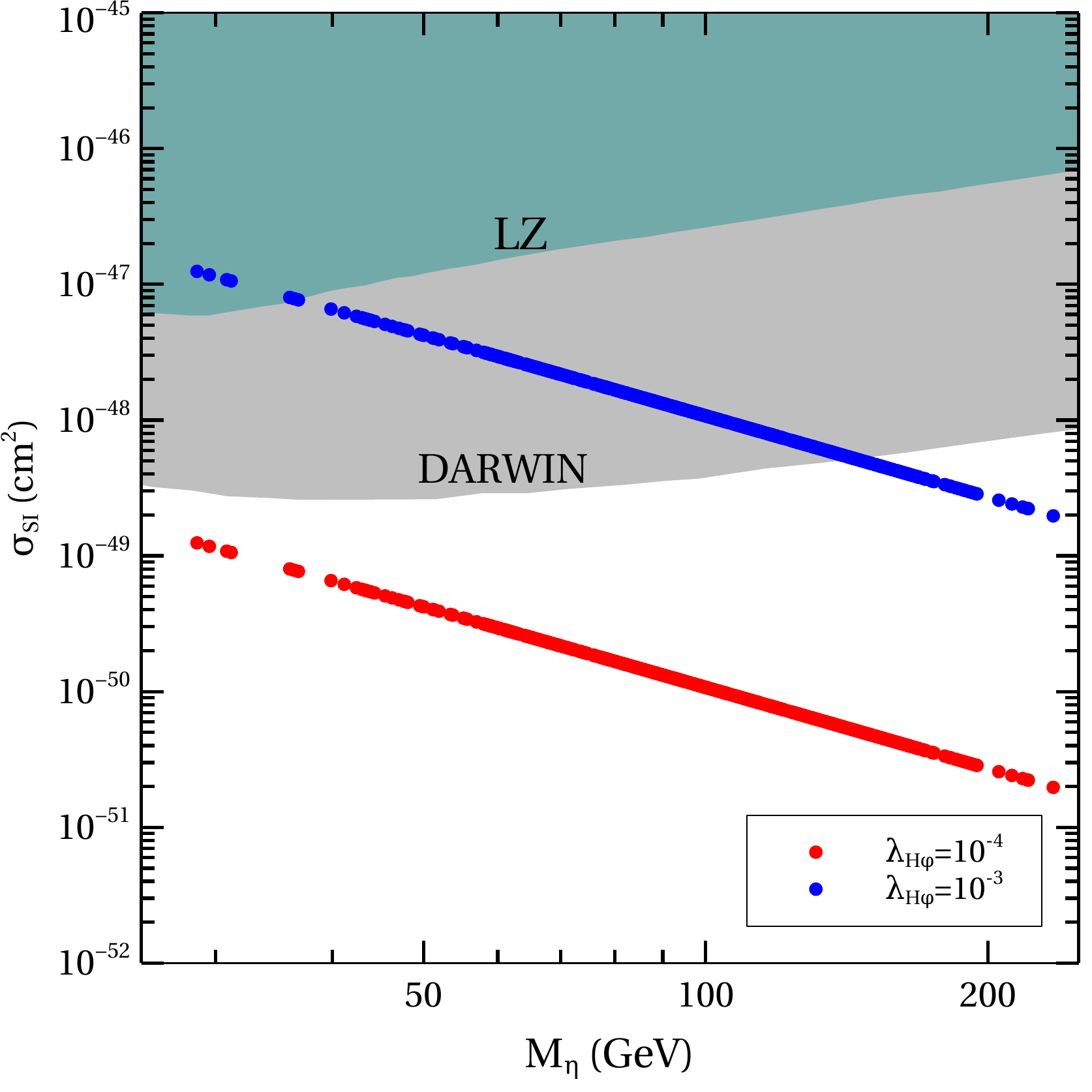}\,
    \caption{Spin-independent direct detection cross-section as a function of DM mass for two different values of Higgs portal coupling $\lambda_{H \phi}$.}
    \label{fig:DD:psi}
\end{figure}

We have chosen Higgs portal coupling $\lambda_{H \phi} = 10^{-4}$ in the above analysis, which is sufficient to bring dark sector in thermal equilibrium with the SM while keeping the direct detection cross-section suppressed. For higher values of Higgs portal coupling, this model can also be tested in the ongoing and future direct detection experiments \cite{LUX-ZEPLIN:2022qhg, DARWIN:2016hyl}. Due to Higgs portal interactions, dominant DM component namely, $\eta$ can scatter off nucleon at tree level leading to spin-independent DM-nucleon scattering cross-section tightly constrained by experiments. In Fig. \ref{fig:DD:psi}, we show the spin-independent DM-nucleon cross-section as a function of $M_\eta$. The current constraint from LZ \cite{LUX-ZEPLIN:2022qhg} and future sensitivity of DARWIN \cite{DARWIN:2016hyl} are shown as shaded regions. The points corresponding to $\lambda_{H \phi}=10^{-4}$ correspond to the data points shown in Fig. \ref{fig:2comp}. Clearly, they remain even out of reach at future experiments. We check that a larger Higgs portal coupling $\lambda_{H \phi}=10^{-3}$ which is still small enough not to alter the relic density analysis discussed above, can bring the direct detection rate within current and future experimental reach for certain range of DM masses, keeping the model verifiable. Semi-annihilating DM can also have interesting indirect detection signatures \cite{Arcadi:2017vis}.

\section{Conclusion}
We have studied a gauge $SU(2)$ symmetry as the origin of dark matter where a scalar quadruplet plays the role of breaking the gauge symmetry spontaneously into a residual $Z_3 \times Z_2$ symmetry responsible for stabilising dark matter. This framework not only gives rise to a fundamental origin of $Z_3$ dark matter, but also predicts a second DM candidate, stabilised by a $Z_2$ symmetry. In this setup, $\phi_I$, the DM matter component stabilised due to the unbroken $Z_2$ symmetry, will always be heavier than the $Z_3$ DM candidate $\eta$ (following Eq. \eqref{eq:mphi}). This also leads to conversion from $\phi_I$ to $\eta$. As a result, the relic density will always be dominated by the lighter $Z_3$ charged component of DM $\eta$.  We have studied the thermal scalar DM scenario considering the dark sector to reach thermal equilibrium with the standard model by virtue of Higgs portal couplings. Dark matter relic abundance is primarily governed by interactions within the dark sector like annihilation, semi-annihilation and conversions with the annihilation into SM particles being sub-dominant due to small Higgs portal couplings required to obey direct detection constraints. Suitable choices of Higgs portal couplings can lead to observable direct detection cross-section keeping the model verifiable. While we have assumed specific mass hierarchy among dark sector particles to study the phenomenology of two scalar DM candidates, other possible mass hierarchies can lead to mixed type of DM consisting of both dark scalar and dark gauge bosons. One can always make $\eta$ to be lighter than the dark gauge bosons ($X_{1,2}$) which are also charged under $Z_3$ symmetry and make them the stable DM candidates. Phenomenology of such dark gauge boson DM has been studied in \cite{Borah:2022phw}. In addition to the possibilities of different thermal DM candidates, one can also have different relic generation mechanism depending upon the relative magnitude of Higgs portal coupling $\lambda_{H\phi}$ and dark sector couplings like $g_D$. For example, if $\lambda_{H\phi}$ is very small but dark sector couplings are sizeable, dark sector may evolve with a different temperature leading to dark freeze-out \cite{Feng:2008mu}. We leave such detailed studies to future works.

\acknowledgements
This work was supported in part by the U.~S.~Department of Energy Grant No. DE-SC0008541. The work of DN is supported by National Research Foundation of Korea (NRF)'s grants with grants No. NRF-2019R1A2C3005009(DN). The work of DB is supported by SERB, Government of India grant MTR/2022/000575. 


\end{document}